\documentclass
[preprintnumbers,nofootinbib,preprint,12pt,showpacs,amsmath] %,amssymb]
{revtex4}

\usepackage{txfonts,bm}

\usepackage{graphicx}

\allowdisplaybreaks[4]

\begin{document}

\title{Coalescence of Rotating Black Holes on Eguchi-Hanson Space }

\author{Ken Matsuno\footnote{E-mail: matsuno@sci.osaka-cu.ac.jp}, 
  Hideki Ishihara\footnote{E-mail: ishihara@sci.osaka-cu.ac.jp}, 
  Masashi Kimura\footnote{E-mail: mkimura@sci.osaka-cu.ac.jp} 
  and 
  Shinya Tomizawa\footnote{E-mail: tomizawa@sci.osaka-cu.ac.jp}
}	

\address{Department of Mathematics and Physics, Graduate School of Science, 
  Osaka City University, 
  3-3-138 Sugimoto, Sumiyoshi-ku, Osaka 558-8585, Japan
}

\begin{abstract}
We obtain new charged rotating multi-black hole solutions 
on the Eguchi-Hanson space in the five-dimensional 
Einstein-Maxwell system with a Chern-Simons term and 
a positive cosmological constant.  
In the two-black holes case, 
these solutions describe the coalescence of 
two rotating black holes 
with the spatial topologies of $\rm S^3$  
into a single rotating black hole 
with the spatial topology of 
the lens space $\rm S^3 / {\mathbb Z}_2$.   
We discuss the differences in   
the horizon areas 
between our solutions and 
the two-centered Klemm-Sabra solutions 
which describe 
the coalescence of two rotating black holes  
with the spatial topologies of $\rm S^3$  
into a single rotating black hole 
with the spatial topology of 
$\rm S^3 $. 

\end{abstract}

\preprint{OCU-PHYS 270}  
\preprint{AP-GR 45}

\date{\today}
\maketitle

\section{Introduction}

Since the advent of the TeV gravity scenarios, i.e., the ADD model~\cite{ADD} and   
the brane world scenario~\cite{RS1,RS2}, 
higher dimensional black objects
have been attracting renewed interest.
One of the reasons is that 
higher dimensional
rotating mini-black holes would be produced
by the collision of protons 
in the Large Hadron Collider (LHC) in these scenarios. 
It would be possible that one detect the Hawking radiation 
from these black holes \cite{BF,GT,IdaOdaPark1,IdaOdaPark2,IdaOdaPark3}. 
Once such black hole productions occur, 
we could expect some of formed rotating black holes to coalesce.
These physical phenomena are expected to give us
new informations on the extra dimensions. Hence, the discovery of 
new higher dimensional black hole solutions would play a crucial role 
in opening a window to extra dimensions.

Higher dimensional black hole solutions has more interesting properties
than the four-dimensional one.
For instance, in the five-dimensional Einstein theory,
there are two types of stationary rotating black hole solutions with
the different horizon topologies, that is,  $\rm \rm S^3$ \cite{MP}
and $\rm S^2 \times \rm S^1$ horizon \cite{ER,MI,PS}.
Both of the solutions asymptote to the five-dimensional Minkowski spacetime
at infinity.   
%Recently, 
Furthermore,   
a lot of asymptotically flat supersymmetric 
black object solutions have been found by various authors.  
The BMPV (Breckenridge, Myers, Peet and Vafa) black hole solutions
\cite{BMPV} 
and the supersymmetric black ring solutions were also found~\cite{EEMR}   
in the five-dimensional $N=2$ supergravity theory   
which 
is one of effective theories 
of the superstring theory and 
contains the five-dimensional Maxwell field
with a Chern-Simons term \cite{sugra}.

Most authors have considered mainly asymptotically flat 
and stationary higher dimensional black hole solutions since
they would be idealized models if such black holes 
are small enough for us to neglect the tension of a brane
or 
effects of compactness of extra dimensions.    
However, if not so, 
we should consider the higher dimensional spacetimes which have 
another asymptotic structure. 
Therefore, it is also important to study black hole solutions 
with a wide class of 
asymptotic structures. 
Recently, the black object solutions with non-trivial 
asymptotic structures 
have been studied by various authors.
For example, Kaluza-Klein black hole solutions  
with squashed $\rm S^3$ horizons   
\cite{DM,GW,IM,TW,SSY,BR} 
asymptote to the three-dimensional flat space 
with a compact twisted $\rm S^1$ fiber at infinity.  
The black ring solutions 
with the same asymptotic structures \cite{BKW,EEMRring,GSYring} were found. 
On the other hand, there exist black object solutions whose spatial infinity 
has the topological structure of lens spaces $L(n;1)=\rm \rm S^3/{\mathbb Z}_n$. 
For instance, 
the solution \cite{IKMT2} represents a pair of non-rotating 
black holes with $\rm S^3 / {\mathbb Z}_2$ infinity. 
It was also found that the supersymmetric black ring solutions 
with the same asymptotic structure~\cite{TIKM}.

There are some dynamical black hole solutions in Einstein-Maxwell theory 
with a positive cosmological constant.
In four-dimensional spacetime, 
Kastor and Traschen found cosmological multi-black hole solutions which
describe the coalescence of charged non-rotating black holes 
by virtue of the positive cosmological constant~\cite{KT,BHKT,BH,NSH}. 
London generalized the Kastor-Traschen solutions 
to higher dimensional ones \cite{LONDON}, which describe   
the coalescence process such that the arbitrary number of non-rotating black holes 
with spherical topology coalesce into a single non-rotating black hole with spherical topology. 
Three of the present authors constructed the different type of black hole solutions 
in the five-dimensional Einstein-Maxwell theory with a positive cosmological constant. 
As shown in Ref.\cite{IKT}, though both solutions also describe 
the coalescence processes of black holes 
by virtue of the existence of the positive cosmological constant, 
the coalescence processes
are essentially different 
in the following point.    
In the five-dimensional Kastor-Traschen solutions, 
two black holes with $\rm \rm S^3$ horizon coalesce into 
a single black hole with $\rm \rm S^3$ horizon, 
while in the solutions in Ref.\cite{IKT},   
two black holes with $\rm \rm S^3$ horizon coalesce 
and convert into a single black hole 
with $\rm \rm S^3 / {\mathbb Z}_2$ horizon on the Eguchi-Hanson space.    
Such the difference arises from the difference in 
the asymptotic structure between both solutions.

Klemm and Sabra also generalized the BMPV solutions \cite{BMPV} to  
the cosmological multi-black hole solutions~\cite{KS}.   
The Klemm-Sabra solutions are also regarded as a generalization of 
the five-dimensional Kastor-Traschen solutions to rotating solutions.    
As will be shown later, 
these solutions describe the coalescence of charged rotating multi-black holes 
with $\rm \rm S^3$ horizon 
into a single rotating black hole 
with $\rm \rm S^3$ horizon  
on the flat space. 
Similarly, we can generalize the non-rotating black hole solutions 
on Eguchi-Hanson space in Ref.\cite{IKT} to rotating black hole solutions 
as the solutions in the five-dimensional Einstein-Maxwell theory with a Chern-Simon term 
and a positive cosmological constant.    
This is the aim of this article.    
We will show 
that this solutions describe the coalescence of two rotating black holes 
with $\rm \rm S^3$ horizon on the Eguchi-Hanson space. 
We will also clarify how the difference in asymptotic structure between our solutions 
and the Klemm-Sabra solutions reflects the difference in coalescence of black holes.

Even if the dynamical properties of 
such solutions are driven   
by the effect of a cosmological constant, the discoveries of 
such black hole solutions are important since it is difficult 
to find exact dynamical black hole solutions in theories 
without a positive cosmological constant, and no one has ever succeeded in 
it as far as we know.  
These black hole solutions are expected that  
they give us the information of dynamical black holes in asymptotic flat spacetimes 
in the case of the sufficiently small cosmological constant.

This article is organized as follows. 
In Section \ref{kaikousei}, 
we give the explicit form of the solutions.  
In Section \ref{reviewKS}, we review the properties of cosmological BMPV solutions 
found by Klemm and Sabra \cite{KS}.  
In Section \ref{hondai}, 
we discuss the coalescence processes of two rotating black holes 
on the Eguchi-Hanson space.   
We compare the coalescence of black holes in our solutions with that 
in the two-centered Klemm-Sabra solutions.  In particular, 
we discuss the difference in the horizon area after the coalescence between both solutions. 
Finally, we give the summary and some discussions.

\section{Solutions} 
\label{kaikousei}

We consider the five-dimensional Einstein-Maxwell system with 
a positive cosmological constant $\Lambda >0$ and a Chern-Simons term.  
The action is given by  
\begin{eqnarray}
 S = \frac{1}{16 \pi G_5} \int d^5 x \sqrt{-g} \left[ 
   R - F_{\mu \nu } F^{\mu \nu } - 4 \Lambda 
   + \frac{2}{3 \sqrt 3} \left( \sqrt{-g} \right)^{-1} 
   \epsilon ^{\mu \nu \rho \sigma \lambda } A_\mu F_{\nu \rho } F_{\sigma \lambda } 
   \right],                                            \label{action}
\end{eqnarray}
where $R$ is the five dimensional scalar curvature, 
$\bm{F} = d \bm{A}$ is the 2-form of the five-dimensional gauge field 
associated with the gauge potential 1-form $\bm{A}$  
and $G_5$ is the five-dimensional Newton constant. 
The action (\ref{action}) with $\Lambda =0$ 
is the bosonic part of the ungauged supersymmetric five-dimensional $N=2$ 
supergravity theory without vector multiplets \cite{sugra}. 

Following this action (\ref{action}), 
we can derive the Einstein equation with 
the positive cosmological constant $\Lambda >0$ 
\begin{eqnarray}
 R_{\mu \nu } -\frac{1}{2} R g_{\mu \nu } + 2\Lambda g_{\mu \nu } 
 = 2 \left( F_{\mu \lambda } F_\nu^{ ~ \lambda } 
  - \frac{1}{4} g_{\mu \nu } F_{\rho \sigma } F^{\rho \sigma } \right), \label{Eineq}
\end{eqnarray}
and the Maxwell equation 
\begin{eqnarray}
 F^{\mu \nu}_{~~~; \nu} + \frac{1}{2 \sqrt 3} \left( \sqrt{-g} \right)^{-1} 
   \epsilon ^{\mu \nu \rho \sigma \lambda } F_{\nu \rho } F_{\sigma \lambda } = 0. 
 \label{Maxeq}  
\end{eqnarray}

We construct new multi-black hole solutions 
on the Eguchi-Hanson base space 
satisfying 
the equations (\ref{Eineq}) and (\ref{Maxeq}). 
The form of the metric and 
the gauge potential 1-form are 
\begin{eqnarray}
 ds^2 &=&- H^{-2} \left[ d\tau + \alpha V^{-1} 
        \left( d\zeta + \bm{\omega} \right) \right]^2 \notag \\ 
        &&+ H \left[ V^{-1} \left( dr^2 + r^2 d\Omega_{S^2}^2 \right) 
                  + V \left( d\zeta + \bm{\omega} \right)^2 \right], \label{EHBH} \\
 \bm{A} &=&\frac{\sqrt 3}{2} H^{-1} 
          \left[ d\tau + \alpha V^{-1} \left( d\zeta + \bm{\omega} \right) \right],
\end{eqnarray}
where $H,~V^{-1}$ and $\bm{\omega}$ are given by 
\begin{eqnarray}
 H &= & \lambda \tau + \frac{M_1}{\left| \bm{r}-\bm{r}_1 \right|}
                  + \frac{M_2}{\left| \bm{r}-\bm{r}_2 \right|}, \label{harmH} \\
 V^{-1} &=&  \frac{N}{\left| \bm{r}-\bm{r}_1 \right|}
            + \frac{N}{\left| \bm{r}-\bm{r}_2 \right|},         \label{harmV} \\
 \bm{\omega} &=&  \left[ \frac{N \left( z-z_1 \right)}{\left| \bm{r}-\bm{r}_1 \right|}
            + \frac{N \left( z-z_2 \right)}{\left| \bm{r}-\bm{r}_2 \right|} \right] d\phi,
                                                                \label{harmome}
\end{eqnarray}
with the constants $M_1,~M_2,~N,~\alpha $  
and $\lambda  = \pm 2 \sqrt{\Lambda /3}$. 
$d\Omega_{S^2}^2 = d\theta ^2 + \sin^2 \theta d\phi ^2$ 
denotes the metric of the unit two-sphere.
The coordinates run the range of $-\infty <\tau <\infty ,~ 0 \le r <\infty ,~ 
0\leq \theta \le \pi ,~ 0\leq \phi \le 2\pi$ and $0\leq \zeta \le 4\pi N$.  
$\bm{r}_i = (x_i,~y_i,~z_i)$ $(i=1,~2)$ 
denote position vectors of the $i$-th nut singularity $N$ 
on the three-dimensional flat space $d{\bm x} \cdot d{\bm x}$. 
The functions $H$ and $V^{-1}$ are 
the solutions of the Laplace equation 
on the three-dimensional flat space.   
The 1-form $\bm{\omega}$ is determined by the equation 
$\nabla \times \bm{\omega} = \nabla ~ V^{-1}$. 

For the appearance of a constant $\lambda $, 
the solution (\ref{EHBH}) is dynamical, i.e., 
it admits no timelike Killing vector field. 
The parameter $\alpha $ in the metric (\ref{EHBH}) 
is an additional parameter for the solution in \cite{IKT}. 
If $\alpha =0$ then the solution (\ref{EHBH}) describes the coalescence of 
two non-rotating black holes on the  Eguchi-Hanson space \cite{IKT}. 
So, we expect that this solution (\ref{EHBH}) describes the coalescence of 
extremely charged two black holes with two equal angular momentums  
on the Eguchi-Hanson space. 
Here and after, we restrict ourselves to considering 
the contracting phase 
with $\lambda  = - 2 \sqrt{\Lambda /3} < 0 $ 
and the range of time $\tau = (-\infty,~ 0)$.

In this article, 
we focus on the regions of the neighborhood of 
$\bm{r} = \bm{r}_i$ $(i=1,~2)$ and
the asymptotic region $r\simeq \infty$
in the solution (\ref{EHBH}).
In the neighborhood of $\bm{r}=\bm{r}_i$, the above metric (\ref{EHBH}) approaches 
to that of 
the Klemm-Sabra solution \cite{KS,CKS}.
Similarly, in the asymptotic region $r\simeq \infty$, 
the local geometry of the metric (\ref{EHBH}) can be regarded 
as that of the Klemm-Sabra solution.
So, in the next section, we review 
the physical properties of the Klemm-Sabra solution.

\section{Review of Klemm-Sabra Solution}
\label{reviewKS}

We review here properties of the Klemm-Sabra solution \cite{KS,CKS}, which 
is the BMPV black hole \cite{BMPV} with a cosmological constant.  
The metric in the cosmological coordinates $(\tau ,~ R)$ are given by 
\begin{eqnarray}
 ds^2 &=&- \left( \lambda \tau + \frac{m}{R^2} \right)^{-2} 
  \left[ d\tau + \frac{j}{2 R^2} \left( d\psi + \cos \theta d\phi \right) \right] ^2 
 \notag \\
        &&+ \left( \lambda \tau + \frac{m}{R^2} \right)
  \left[ dR^2 + \frac{R^2}{4} \left\{ d\Omega_{\rm S^2}^2 
  +\left( d\psi + \cos \theta d\phi \right)^2 \right\} \right], \label{KSmet}
\end{eqnarray}
where 
$m$ and $j$ are constants which
specify 
the mass
 and the angular momentum. 
The curvature singularity exist at $\lambda \tau R^2=-m$.   
Indeed, setting $\tau $ to be $\tau + \lambda ^{-1}$ 
and taking the limit $\lambda \to 0$,
we find   
the metric (\ref{KSmet}) reduces to the BMPV black hole solution \cite{BMPV}.

One obtains
the expansions $\theta _\pm$ of the outgoing and ingoing null geodesics 
for the $\tau=$ const. and $R=$ const. 
surface as 
\begin{eqnarray}
 \theta _\pm =\lambda \pm \frac{2x}{\sqrt{( x+m ) ^3 -j^2}},  
\end{eqnarray}
where we introduced a coordinate $x = \lambda \tau R^2$. 
Thus, the horizon occur at $x$ such that 
\begin{eqnarray}
 \lambda ^2 \left[ \left( x+m \right)^3 -j^2 \right] -4x^2 =0. \label{difofhors}
\end{eqnarray}

The solution \eqref{KSmet} seems to be dynamical for the dependence of $\tau $, 
but it is stationary.  
Actually, one can introduce stationary coordinates $(\hat t,~\hat r,~\hat \psi)$ 
for the solution (\ref{KSmet}) as follows, 
\begin{eqnarray}
 \lambda \tau R^2 = \hat r^2 -m, \quad  
 \left( \lambda \tau \right)^{-1} d\tau 
 = d\hat t + \hat f(\hat r) d\hat r, 
 \quad 
 d\psi = d\hat \psi + \hat h(\hat r) d\hat r,  \label{KSstatct}
\end{eqnarray}  
with
\begin{eqnarray}
 \hat f(\hat r) 
 = \frac{2\lambda \hat r \left( \hat r^6 -j^2 \right) / \left( \hat r^2 -m \right)}
 {\lambda ^2 \left( \hat r^6 -j^2 \right) -4 \left( \hat r^2 -m \right)^2}, \quad
 \hat h(\hat r) = \frac{4 \lambda j \hat r}
 {\lambda ^2 \left( \hat r^6 -j^2 \right) -4 \left( \hat r^2 -m \right)^2}. 
\end{eqnarray} 
The form of the metric (\ref{KSmet}) after the above coordinates transformation 
now becomes 
\begin{eqnarray}
 ds^2 &=& \frac{\lambda ^2}{4} \hat r^2 d\hat t^2 - U^2 (\hat r) 
 \left[ d\hat t + \frac{j}{2 \hat r^2} U^{-1}(\hat r) \left( d\hat \psi + 
 \cos \theta d\phi \right) \right]^2
 \notag \\
 && +\frac{d\hat r^2}{W(\hat r)} + \frac{\hat r^2}{4} \left[ d\Omega_{\rm S^2}^2 
  +\left( d\hat \psi + \cos \theta d\phi \right)^2 \right], \label{CKSmet}
\end{eqnarray}
where the functions $U(\hat r)$ and $W(\hat r)$ are 
\begin{eqnarray}
 U(\hat r) = 1 - \frac{m}{\hat r^2}, \quad
 W(\hat r) = \left( 1 - \frac{m}{\hat r^2} \right) ^2 - \frac{\lambda ^2}{4} \hat r^2 
         + \frac{\lambda ^2 j^2}{4\hat r^4}. \label{defhorizon}
\end{eqnarray}
From (\ref{difofhors}), (\ref{KSstatct}) and (\ref{defhorizon}),   
we see that the horizon occur at $\hat r$ such that $W(\hat r)=0$.

\begin{figure}[htbp]
 \begin{center}
 \includegraphics[width=10cm,clip]{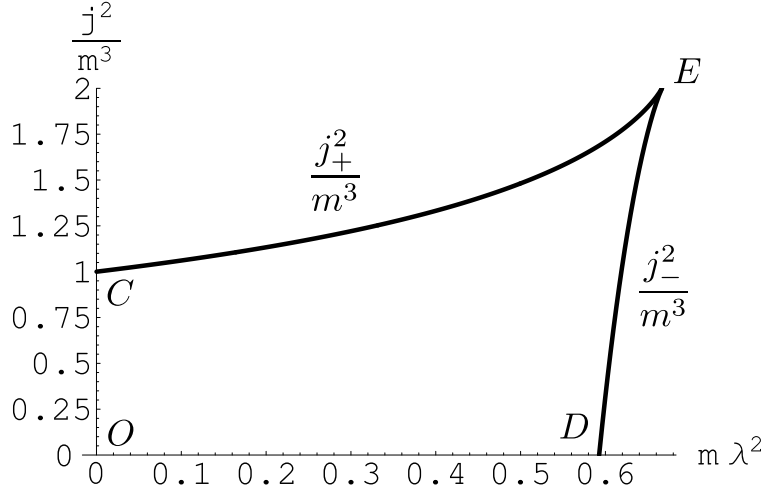}
 \end{center}
 \caption{
This figure shows the region of parameters such that the solutions
have no naked singularity.
The vertical axis and the horizontal axis denote $j^2/m^3$ and $m\lambda^2$, 
respectively.
The curves $CE$ and $DE$ correspond to
$j^2/m^3=j^2_+/m^3$ and $j^2/m^3= j^2_-/m^3$, respectively.
The solutions lying in the region $ODEC$ have three horizons.
On $CE$ the black hole horizon $x_+$ coincides with  
the inner horizon $x_-$
and on $DE$ the black hole horizon $x_+$ coincides with  
the cosmological horizon $x_c$.
Outside the region $ODEC$ 
there exist naked singularities.
}
 \label{hannizu}
\end{figure}

The equation (\ref{difofhors}) has three real roots 
$x_-,~x_+,~x_c$ ($x_- \le 0 \le x_+ \le x_c$),
where $x_-,~x_+,~x_c$ correspond to the inner horizon, the black hole horizon
and the cosmological horizon, respectively,
if the mass parameter $m$ and the angular momentum parameter $j$ 
satisfies the following conditions, 
\begin{eqnarray}
 0\le m \lambda^2 \le \frac{2}{3}, \quad 
 j_-^2 (m) \le j^2 \le j_+^2 (m), \label{hanni}
\end{eqnarray}
where  
\begin{eqnarray} 
 j_\pm^2 (m) = \frac{4}{27 \lambda ^6} \left[9 m\lambda ^2 (8-3 m\lambda ^2) -32 \pm 8
   \sqrt{2} (2-3 m\lambda ^2 )^{3/2}\right]. 
\end{eqnarray}
In the case of $j=j_+$, the black hole horizon $x_+$ coincides with  
the inner horizon $x_-$, 
and in the case of $j=j_-$, the black hole horizon $x_+$ coincides with  
the cosmological horizon $x_c$.  
The naked singularity appears if $m$ and $j$ are out of the ranges (\ref{hanni}).
We draw the region of $(m,~j)$ satisfying the condition (\ref{hanni})
in FIG.\ref{hannizu}.
Next, we focus on the conditions for the absence of closed timelike curves (CTCs) 
outside the black hole horizon $x_+(m,~j)$. 
These CTCs occur if and only if 
the two dimensional $(\psi ,~ \phi )$ part of the metric (\ref{KSmet}), 
namely,   
$g_{\rm 2D}$    
has a negative eigenvalue. 
We must check the condition 
$g_{\psi \psi } >0$ and det $g_{\rm 2D} >0$ for $x> x_+ >0$. 
In this case, explicit forms of these components are given by 
\begin{eqnarray} 
  g_{\psi \psi } = \frac{( x+m )^3 -j^2}{4 (x + m) ^2}, \quad 
  {\rm det} ~g_{\rm 2D} = 
   \frac{( x+m )^3 -j^2}{16 (x + m)} \sin ^2 \theta.  
\end{eqnarray}
Since the numerators of $g_{\psi \psi }$ and det $g_{\rm 2D}$ are 
monotonically increasing functions of $x$, 
it is sufficient to show 
$g_{\psi \psi } >0$ and det $g_{\rm 2D} >0$ on the horizon $x_+$. 
Actually, we see that 
\begin{eqnarray} 
  g_{\psi \psi } = \left[ \frac{x_+}{\lambda (x_+ +m)} \right]^2 >0, \quad 
  {\rm det} ~g_{\rm 2D} = 
   \frac{x_+ ^2}{4\lambda^2 (x_+ +m)} \sin ^2 \theta >0, 
\end{eqnarray}
for $x_+>0$ and $m > 0$. 
Fortunately, we obtain the regular black hole solutions 
with parameters $(m,~j)$ satisfying the condition (\ref{hanni}) 
which have no CTCs outside the black hole horizon.  

The induced metric 
on the black hole horizon $x=x_+(m,~j)$ becomes 
\begin{eqnarray} 
  ds^2 _H = \frac{x_+ + m}{4} 
  \left[ d\Omega_{S^2}^2 
  + \frac{(x_+ + m)^3 -j^2}{(x_+ + m)^3} 
  \left( d\psi + \cos \theta d\phi \right)^2 \right], \label{horindmet}
\end{eqnarray} 
which implies the shape of horizon is the squashed $\rm S^3$, 
a twisted $\rm S^1$ fiber bundle over an $\rm S^2$ base space 
with the different sizes, 
for the presence of angular momentum parameter.

From (\ref{difofhors}) and (\ref{horindmet})
we obtain the expression of the area of the black hole horizon $x=x_+(m,~j)$ 
as 
\begin{eqnarray} 
  A _H = \frac{2}{\lambda }x_+ (m,~j)  A_{\rm S^3}, \label{horarea}
\end{eqnarray}
where $A_{\rm S^3}$ denotes the area of the unit $\rm S^3$.
We will see the change in the horizon area before and after coalescence 
with this expression (\ref{horarea}).

\section{Coalescence of two Rotating Black Holes } 
\label{hondai}

\subsection{Asymptotic Behavior of Black Holes at Early Time and Late Time}

First, 
we investigate the asymptotic behaviors
of the metric (\ref{EHBH}) in the neighborhood of ${\bm r}={\bm r}_i$ $(i=1,~2)$.  
In this region, the metric (\ref{EHBH}) takes the form of  
\begin{eqnarray}
 ds^2 &\simeq &- \left( \lambda \tau + \frac{m_i}{\tilde r^2} \right)^{-2} 
  \left[ d\tau + \frac{j}{2 \tilde r^2} \left( d\psi + \cos \theta d\phi \right) \right]^2 
 \notag \\
        &&+ \left( \lambda \tau + \frac{m_i}{\tilde r^2} \right)
  \left[ d\tilde r^2 + \frac{\tilde r^2}{4} \left\{ d\Omega_{S^2}^2 
  +\left( d\psi + \cos \theta d\phi \right)^2 \right\} \right], \label{ETBHhon}
\end{eqnarray}
where we introduced the coordinates  
$\tilde r^2 = 4N r,~ \psi = \zeta / N,~ m_i = 4N M_i$ and $j = 8\alpha N^3$. 
This metric is 
equal to
that of
the Klemm-Sabra solutions (\ref{KSmet}) 
with the mass parameters $m_i$ and the angular momentum parameter $j$.
As discussed in the previous section \ref{reviewKS}, 
this solution (\ref{ETBHhon}) admits three horizons at $x=x_\pm,~x_c$,  
in the coordinate $x = \lambda \tau \tilde r^2$. At the early time $\tau \simeq -\infty$, 
sufficiently small squashed $\rm S^3$ spheres 
centered at $\bm r = \bm r _i$ are always outer trapped
since there are solutions for $\theta _+ =0$ 
at $\tilde r^2 = x_+(m_i,~j) / (\lambda \tau )$.
Thus, at the early time, there are two rotating black holes with 
the horizon topology $\rm S^3$.

Next, we focus on the asymptotic region of the solution (\ref{EHBH}), 
$r \simeq \infty $. 
We assume the separation of two black holes 
$\left| \bm{r}_1 - \bm{r}_2 \right|$ is much smaller than $r$. 
In this region, the metric (\ref{EHBH}) behaves as  
\begin{eqnarray}
 ds^2 &\simeq &- \left[ \lambda \tau + \frac{2 (m_1 +m_2)}{\rho^2} \right]^{-2} 
  \left[ d\tau + \frac{8j}{2 \rho^2} \left( \frac{d\psi}{2} + \cos \theta d\phi \right) \right]^2 \notag \\
        &&+ \left[ \lambda \tau + \frac{2 (m_1 +m_2)}{\rho^2} \right]
  \left[ d\rho^2 + \frac{\rho^2}{4} \left\{ d\Omega_{S^2}^2 
  +\left( \frac{d\psi}{2} + \cos \theta d\phi \right)^2 \right\} \right],   
  \label{LTBHhon}
\end{eqnarray}
where we introduced the coordinates $\rho^2 = 8N r,~\psi = \zeta / N$ 
and parameters $m_i = 4N M_i$ and $j = 8\alpha N^3$, as same as in (\ref{ETBHhon}). 
This metric (\ref{LTBHhon}) resembles that of the Klemm-Sabra solution (\ref{KSmet}) 
with the mass parameter $2 (m_1 +m_2)$ and angular momentum parameter $8j$. 

Like the Klemm-Sabra solution (\ref{KSmet}), at the late time $\tau \simeq 0$,  sufficiently large squashed $\rm S^3$ sphere becomes outer trapped, 
since $\theta _+ =0$ at 
$\rho ^2 = x_+ \left( 2 (m_1 +m_2),~8j \right) / (\lambda \tau )$, 
which give an approximately large sphere. 
However, we see this solution (\ref{LTBHhon}) differs from 
the Klemm-Sabra solution (\ref{KSmet}) in the following point; 
each $\rho =$ const. surface in the $\tau =$ const. hypersurface 
of the metric (\ref{LTBHhon})
denotes topologically the lens space $\rm S^3 /{\mathbb Z}_2$, 
while in the Klemm-Sabra solution (\ref{KSmet}), 
it is diffeomorphic to $\rm S^3$. 
The difference between these metrics appears 
in (\ref{ETBHhon}) and (\ref{LTBHhon}): 
a term $d\psi $ in $\rm S^3$ metric (\ref{ETBHhon}) is replaced by 
a term $d\psi /2$ in $\rm S^3 / {\mathbb Z}_2$ metric (\ref{LTBHhon}). 
Therefore, at the late time $\tau \simeq 0$, the topology of 
the outer trapped surface is the lens space $\rm S^3 / {\mathbb Z}_2$. 

Hence, from these results, we find that the solution (\ref{EHBH}) 
describes the dynamical situation such that two rotating black holes 
with the spatial topologies of $\rm S^3$ coalesce and convert into 
a single rotating black hole 
with the spatial topology of the lens space $\rm S^3 / {\mathbb Z}_2$. 
Thus, 
at the early time, 
there are two rotating black holes 
specified by $(m_1,~j)$ and $(m_2,~j)$. 
At the late time, 
there is a single rotating black hole 
specified by $(2 (m_1 +m_2),~j)$. 
Here and after, 
we call such relations 
\lq\lq mapping rule\rq\rq.

\subsection{Typical Processes in Klemm-Sabra solutions}

We compare the coalescence processes described by 
our solutions (\ref{EHBH}) 
with the coalescence of two rotating black holes with 
$S^3$ horizon 
into a single rotating black hole with 
$S^3$ horizon.    
For this purpose,  
let us extend a single Klemm-Sabra solution (\ref{KSmet}) to  
the two-centered Klemm-Sabra solution 
which denotes 
two rotating black holes 
with mass parameters $m_1,~m_2$ and 
angular momentum parameters $j_1,~j_2$ 
at the early time, 
\begin{eqnarray}
ds^2 = - H^{-2} \left( d\tau + dx^a J_a^{~b} \partial _b K \right)^2 
       + H d\bm{x} \cdot d\bm{x}, \label{2CKS}
\end{eqnarray}
with 
\begin{eqnarray}
 H &=&  
  \lambda \tau + \frac{ m_1 }{ \left| \bm{x}-\bm{x}_1 \right|^2 }
  + \frac{ m_2 }{ \left| \bm{x}-\bm{x}_2 \right|^2 }, \\
 K &=& 
  \frac{ j_1 /2 }{ \left| \bm{x}-\bm{x}_1 \right|^2 }
  + \frac{ j_2 /2 }{ \left| \bm{x}-\bm{x}_2 \right|^2 }, 
\end{eqnarray}
where $J$ is a complex structure,  
$\bm{x} = (x,~y,~z,~w)$, 
$\bm{x}_i$ $(i=1,~2)$ are position vectors in ${\mathbb E}^4$ and  
$m_i$, $j_i$ 
are positive constants.  
The \lq\lq mapping rule"  for this solution (\ref{2CKS}) 
becomes as follows; 
At the early time, 
there are two rotating black holes 
specified by $(m_1,~j_1)$ and $(m_2,~j_2)$. 
At the late time, 
there is a single rotating black hole 
specified by $(m_1 + m_2,~j_1 + j_2)$.

\begin{figure}[htbp]
 \begin{center}
 \includegraphics[width=10cm,clip]{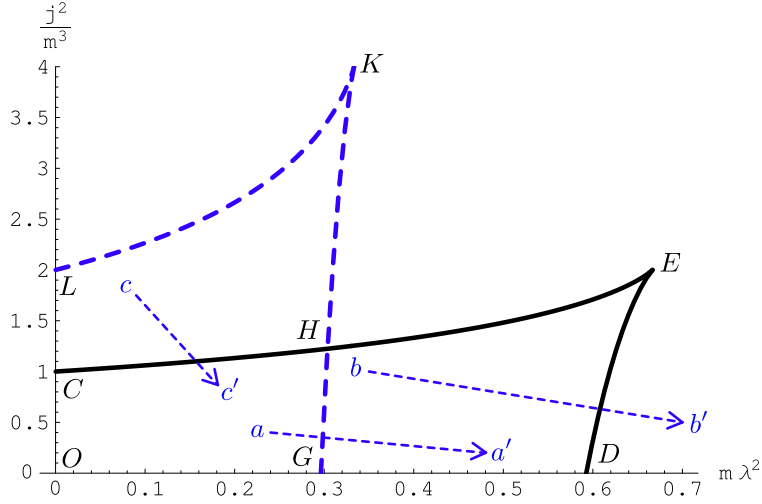}
 \end{center}
 \caption{
 This figure shows typical processes 
 described by the two-centered Klemm-Sabra solutions (\ref{2CKS}). } 
 \label{ehgattai3}
\end{figure}

Here, to compare our solution (\ref{EHBH}) 
with this solution (\ref{2CKS}), 
we restrict ourselves to the solution (\ref{2CKS}) 
with the same mass parameters $m = m_1 = m_2$ and 
the same angular momentum parameters $j = j_1 = j_2$. 
According to this \lq\lq mapping rule\rq\rq, 
we discuss types of process 
by using dimensionless parameters 
$m\lambda ^2$ and $j^2 / m^3$. 
These parameters 
are mapped as
$(m\lambda ^2,~j^2 / m^3) \to \left( 2m\lambda ^2,~(j^2 / m^3) /2 \right)$ 
(see FIG.\ref{ehgattai3}). 

Any solutions  
lying in the region $ODEC$ 
describe regular initial condition 
such that 
there exist two isolated apparent horizons. 
In contrast,  
according to the above \lq\lq mapping rule\rq\rq, 
any solutions
lying in the region $OGKL$ 
describe a single rotating black hole with $\rm \rm S^3$ horizon 
at the late time. 
So, any solutions lying in the region $OGHC$ 
describe a coalescence of two rotating black holes. 
There are four types of regions, namely,  
$OGHC,~ GDEH,~ CHKL$ and outside of $DEHKL$.  
These regions correspond to the four kinds of process.
The blue dashed arrows represent typical processes.

The process $a \to a^\prime$%$\overrightarrow{aa'}$  
describes the situation such that 
two rotating black holes with $\rm \rm S^3$ horizon 
coalesce and convert into a single rotating black hole 
with $\rm \rm S^3$ horizon. 
The arrow $b \to b^\prime$%$\overrightarrow{bb'}$
describes 
the situation such that 
there are two isolated apparent horizons at the early time,
and there exist a naked singularity at the late time. 
The process $c \to c^\prime$%$\overrightarrow{cc'}$
describes 
the situation such that 
there is not an apparent horizon and CTCs appear at the early time,  
and at the late time, 
there exist   
a single rotating black hole with $\rm \rm S^3$ horizon 
and there is no CTC outside the horizon.

\subsection{Typical Processes in our solutions}

\begin{figure}[htbp]
 \begin{center}
 \includegraphics[width=10cm,clip]{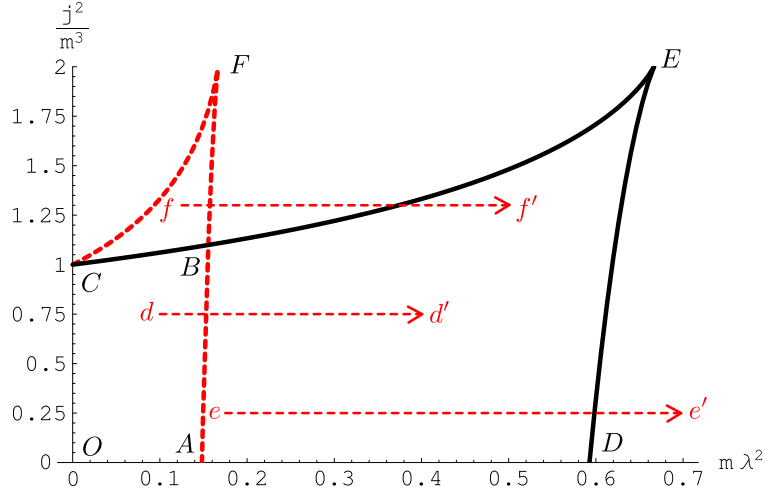}
 \end{center}
 \caption{
 This figure shows typical processes 
 described by our solutions (\ref{EHBH}). }
 \label{ehgattai2}
\end{figure}

Our solutions (\ref{EHBH}) also describe similar processes to those
described by two center Klemm-Sabra solution (\ref{2CKS}).
Now, 
we compare our cases with two-centered Klemm-Sabra's cases.
We restrict ourselves to the solution (\ref{EHBH})
with the same mass parameters $m = m_1 = m_2$. 
According to the \lq\lq mapping rule"  of our solutions (\ref{EHBH}),
the dimensionless parameters 
$m\lambda ^2$ and $j^2 / m^3$
are mapped as
$(m\lambda ^2,~j^2 / m^3) \to \left( 4m\lambda ^2,~j^2 / m^3 \right)$ 
(see FIG.\ref{ehgattai2}).

As shown in FIG.\ref{ehgattai3}, 
any solutions 
lying in the region $ODEC$ 
describe regular initial condition 
such that 
there exist two isolated apparent horizons. 
In contrast,  
according to the \lq\lq mapping rule"  of our solution (\ref{EHBH}), 
any solutions 
lying in the region $OAFC$ 
describe a single rotating black hole 
with $\rm \rm S^3 / {\mathbb Z}_2$ horizon 
at the late time. 
So, any solutions lying in the region $OABC$ 
describe a coalescence of two rotating black holes. 
There are four types of regions, namely,  
$OABC,~ ADEB,~ CBF$ and outside of $DEBFC$.  
These regions correspond to the four kinds of process.
The red dashed arrows represent typical processes.

The process $d \to d^\prime$%$\overrightarrow{dd'}$  
describes the situation such that 
two rotating black holes with $\rm \rm S^3$ horizon 
coalesce and convert into a single rotating black hole 
with $\rm \rm S^3 / {\mathbb Z}_2$ horizon. 
The arrow $e \to e^\prime$%$\overrightarrow{ee'}$
describes 
the situation such that 
there are two isolated apparent horizons at the early time, 
and there exist a naked singularity at the late time. 
The process $f \to f^\prime$%$\overrightarrow{ff'}$
describes 
the situation such that 
there is not an apparent horizon but CTCs at the early time,  
while at the late time, 
there exist   
a single rotating black hole with $\rm \rm S^3 / {\mathbb Z}_2$ horizon 
and there is no CTC outside the horizon.

\begin{figure}[htbp]
 \begin{center}
 \includegraphics[width=10cm,clip]{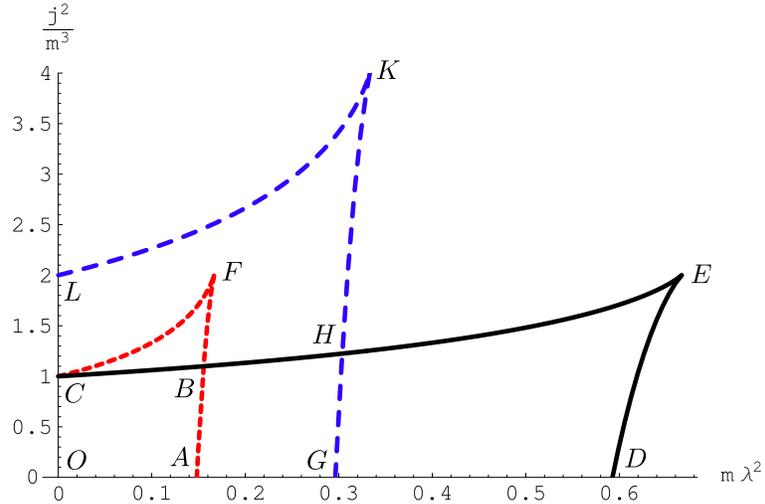}
 \end{center}
\caption{
 This figure shows the superposition of FIG.\ref{ehgattai3} and FIG.\ref{ehgattai2}.
 } 
 \label{3regions}
\end{figure}

From above discussions, there is 
the featuring difference in   
\lq\lq mapping rule" between our solution (\ref{EHBH}) and
two-centered Klemm-Sabra solution (\ref{2CKS})
in the region $BHKLCF$ in FIG.\ref{3regions}.  
At the early time, 
both solutions in this region 
have no apparent horizon. 
At the late time, 
the two-centered Klemm-Sabra solution (\ref{2CKS}) describes 
a single rotating black hole with $\rm \rm S^3$ horizon 
while our solution (\ref{EHBH}) describes
a naked singularity. 

\subsection{Comparison of Horizon Areas}
We compare the area of a single rotating black hole 
formed by the coalescence of two rotating black holes at the late time. 
We assume that 
each black hole in our solution (\ref{EHBH})
has the same mass, angular momentum and horizon area 
as that in the two-centered Klemm-Sabra solution at the early time.
Then, from the equation (\ref{horarea}), 
the horizon areas in Klemm-Sabra solutions and our solutions
at the early time, $A_{\rm Flat}^{(e)}$ and $A_{\rm EH} ^{(e)}$, are given by 
\begin{eqnarray}
 A_{\rm Flat} ^{(e)} = A_{\rm EH} ^{(e)} = 
 2 \times \frac{2}{\lambda } x_+ (m,~j) A_{\rm \rm S^3}.  
\end{eqnarray}  

On the other hand, according to the \lq\lq mapping rules" of both solutions, 
the horizon areas  
at the late time, $A_{\rm Flat} ^{(l)}$ and $A_{\rm EH} ^{(l)}$, are given by  
\begin{eqnarray}
 A_{\rm Flat} ^{(l)} &=& \frac{2}{\lambda } x_+ (2m,~2j) A_{\rm \rm S^3},  
 \label{areaFlat} \\
 A_{\rm EH} ^{(l)} &=& \frac{2}{\lambda } x_+ (4m,~8j) \frac{A_{\rm \rm S^3}}{2},  
 \label{areaEH} 
\end{eqnarray}
respectively. 
Note the factor $1 / 2$ in the equation (\ref{areaEH})
reflects the fact that the black hole at the late time 
after coalescence of two black holes is 
topologically the lens space $\rm \rm S^3 / {\mathbb Z}_2$.

\begin{figure}[htbp]
 \begin{center}
 \includegraphics[width=10cm,clip]{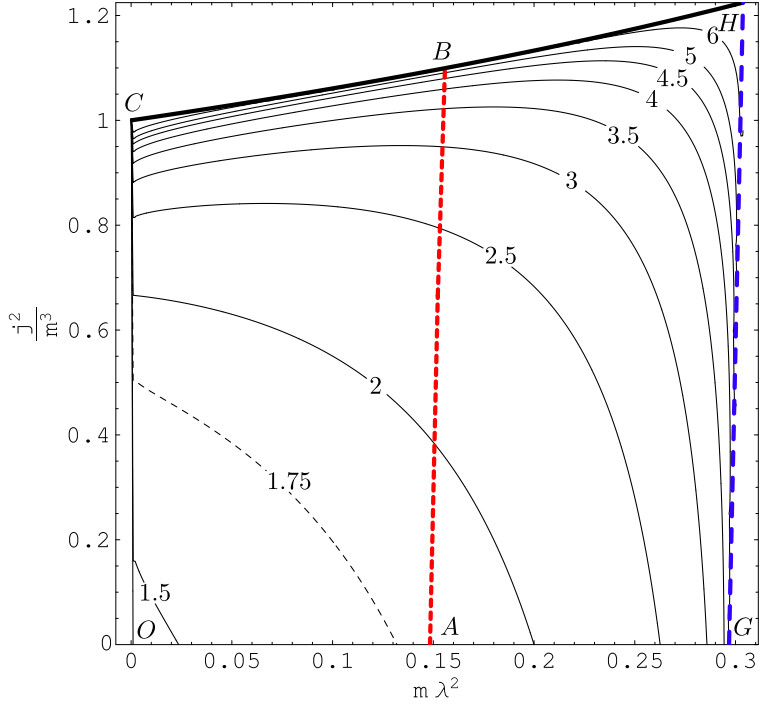}
 \end{center}
\caption{
 This figure shows the dependence of the ratio    
 $\left. A_{\rm Flat} ^{(l)}  \right/ A_{\rm Flat} ^{(e)}$ 
 on $m \lambda ^2$ (horizontal axis) and $j^2 / m^3$ (vertical axis).  
 The curves in this figure denote 
 $\left. A_{\rm Flat} ^{(l)}  \right/ A_{\rm Flat} ^{(e)}=$ const.  
 } 
 \label{areahenkaflat}
\end{figure}

\begin{figure}[htbp]
 \begin{center}
 \includegraphics[width=10cm,clip]{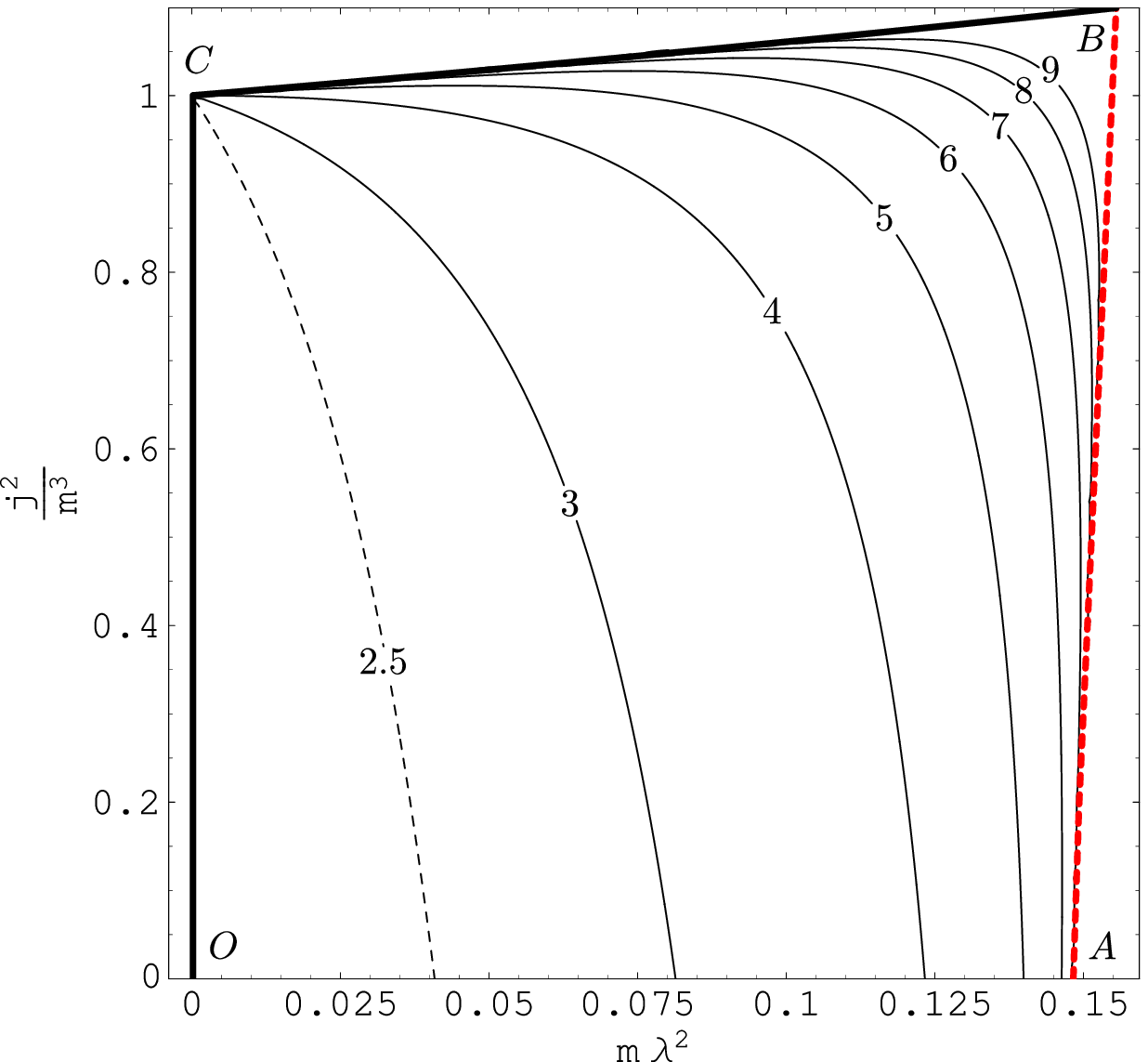}
 \end{center}
\caption{
 This figure shows the dependence of the ratio    
 $\left. A_{\rm EH} ^{(l)}  \right/ A_{\rm EH} ^{(e)}$ 
 on $m \lambda ^2$ (horizontal axis) and $j^2 / m^3$ (vertical axis).  
 The curves in this figure denote 
 $\left. A_{\rm EH} ^{(l)}  \right/ A_{\rm EH} ^{(e)}=$ const.    
 } 
 \label{areahenkaeh}
\end{figure}

Now, we consider the ratio of horizon areas at the early time to at the late time 
$\left. A ^{(l)}  \right/ A ^{(e)}$ 
in both solutions. 
The dependence of the ratio on $(m\lambda^2,j^2/m^3)$ in Klemm-Sabra solution
is shown in FIG.\ref{areahenkaflat}.
The same in our solution is shown in  FIG.\ref{areahenkaeh}. In all regions, 
$\left. A_{\rm Flat} ^{(l)}  \right/ A_{\rm Flat} ^{(e)} > 1$ and 
$\left. A_{\rm EH} ^{(l)}  \right/ A_{\rm EH} ^{(e)} > 1$.
This means that the horizon areas increase by the coalescence.
Qualitative behavior of the ratio near the boundary
$GH$ in FIG.\ref{areahenkaflat} is similar 
to that near the boundary $AB$ in FIG.\ref{areahenkaeh}.

However, the behaviors of the ratio near $OC$ are different.
Here, we focus on the behaviors in $\lambda \to 0$ limit.
From FIG.\ref{areahenkaflat}, the ratio 
$\left. \left. A_{\rm Flat} ^{(l)}  \right/ A_{\rm Flat} ^{(e)} \right| _{\lambda \to 0}
= \sqrt{(2m^3 - j^2) / (m^3 - j^2)}$
depends on the ratio $j^2 / m^3$ along the line $OC$.
In contrast,
from FIG.\ref{areahenkaeh},
the ratio of horizon is independent of the ratio $j^2/m^3$,
that is, 
$\left. \left. A_{\rm EH} ^{(l)}  \right/ A_{\rm EH} ^{(e)} \right| _{\lambda \to 0}
= 2$ 
on the line $OC$. 

\begin{figure}[htbp]
 \begin{center}
 \includegraphics[width=10cm,clip]{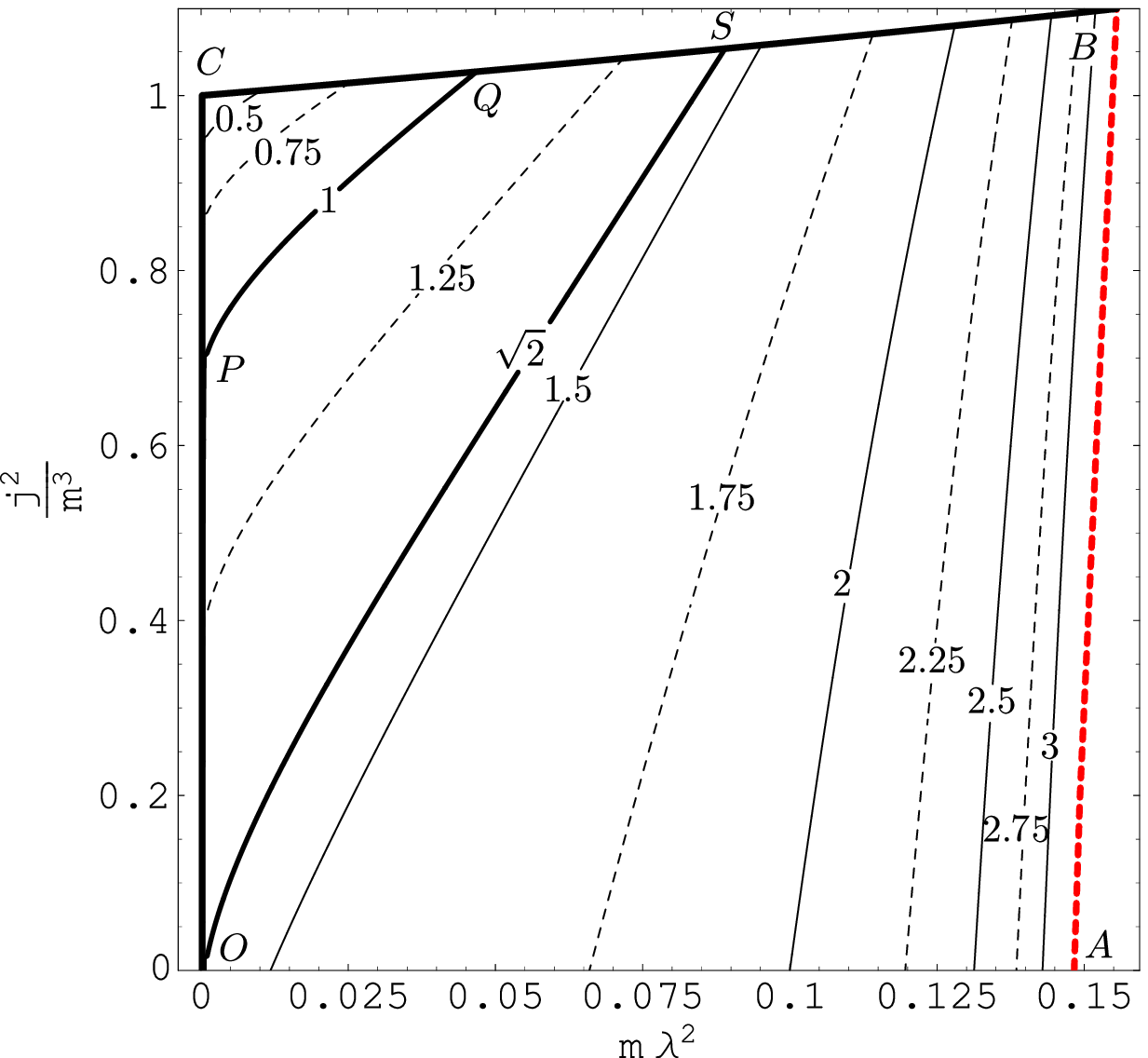}
 \end{center}
\caption{
 This figure shows the dependence of the ratio    
 $\left. A_{\rm EH} ^{(l)}  \right/ A_{\rm Flat} ^{(l)}$ 
 on $m \lambda ^2$ (horizontal axis) and $j^2 / m^3$ (vertical axis).  
 The curves in this figure denote 
 $\left. A_{\rm EH} ^{(l)}  \right/ A_{\rm Flat} ^{(l)}=$ const. 
 Here, 
 the ratio of horizon area becomes  
 $ 0 < \left. A_{\rm EH} ^{(l)}  \right/ A_{\rm Flat} ^{(l)} < \sqrt 2 $ 
 in the region $OSC$. 
 This behavior is one of the unique properties of the solution (\ref{EHBH}) 
 with a presence of rotations. } 
 \label{areahenka}
\end{figure}

\begin{figure}[htbp]
 \begin{center}
 \includegraphics[width=10cm,clip]{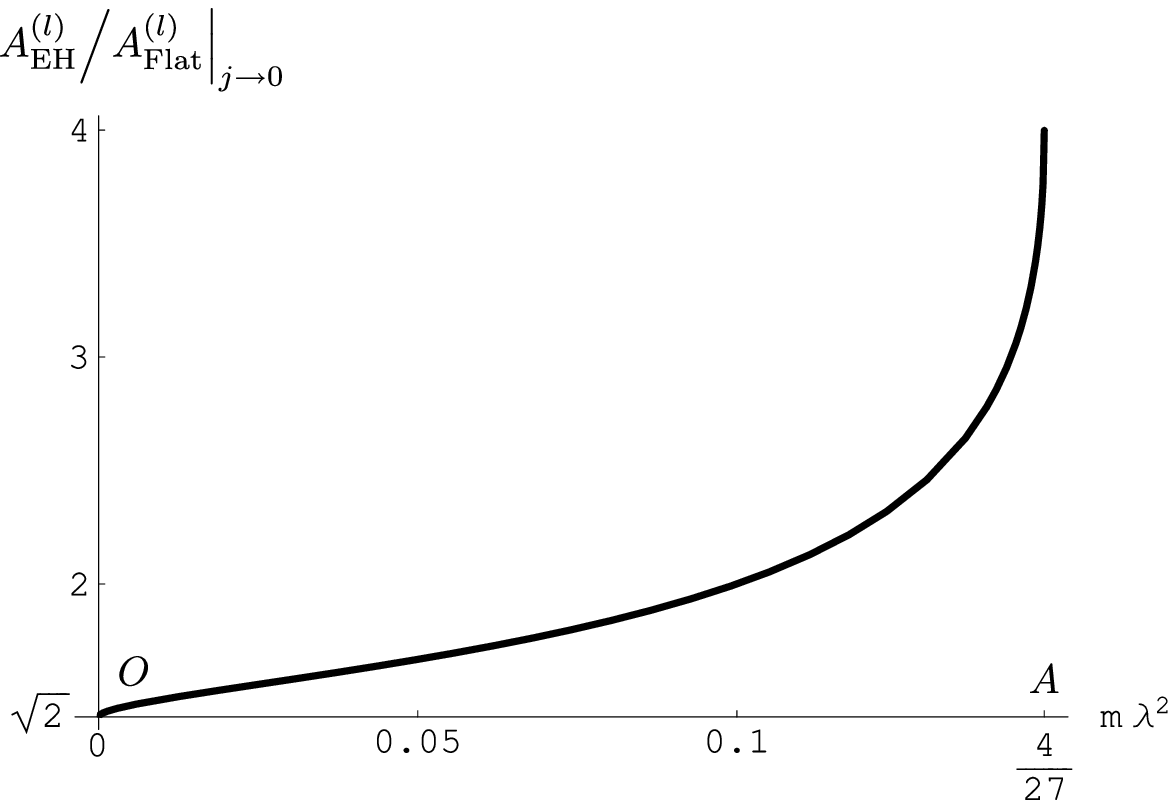}
 \end{center}
 \caption{
 This figure shows the dependence of 
 $\left. \left. A_{\rm EH} ^{(l)}  \right/ A_{\rm Flat} ^{(l)} \right| _{j \to 0}$  
 on $m \lambda ^2$ on the line $OA$. 
 We see that 
 $\left. \left. A_{\rm EH} ^{(l)}  \right/ A_{\rm Flat} ^{(l)} \right| _{j \to 0}$ 
 is  
 as same as in the non-rotating case \cite{IKT}.  }
 \label{areadannmenoa}
\end{figure}

\begin{figure}[htbp]
 \begin{center}
 \includegraphics[width=10cm,clip]{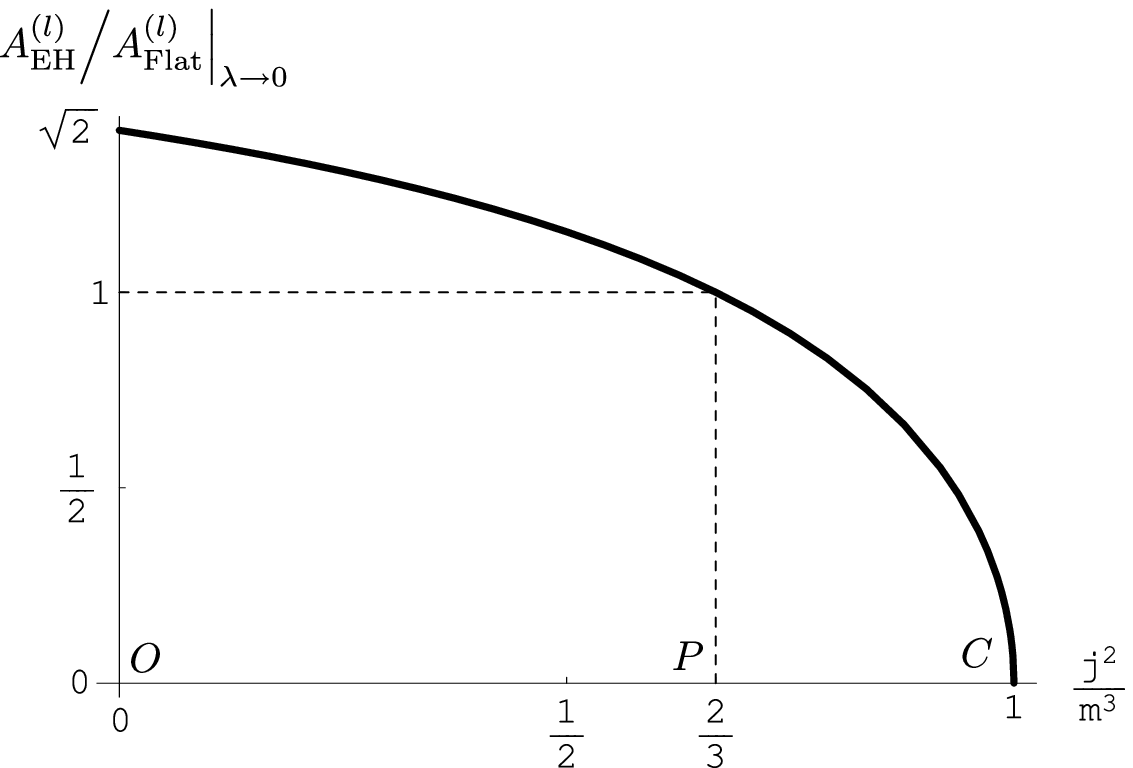}
 \end{center}
 \caption{
 This figure shows that the dependence of 
 $\left. \left. A_{\rm EH} ^{(l)}  \right/ A_{\rm Flat} ^{(l)} \right| _{\lambda \to 0}$  
 on $j^2 /m^3$ on the line $OC$. 
 For large angular momentum, i.e., 
 $2/3 < j^2 / m^3 < 1$, 
 the area of black hole horizon after coalescence 
 on the Eguchi-Hanson space 
 is smaller than that on the flat space, that is,  
 $0 < \left. A_{\rm EH} ^{(l)}  \right/ A_{\rm Flat} ^{(l)} < 1 $. 
 }
 \label{areadannmenoc}
\end{figure}

In turn, to clarify the differences in 
the ratio of the horizon areas of 
two-centered Klemm-Sabra solution to that of our solution, 
we consider the ratio 
$\left. A_{\rm EH} ^{(l)}  \right/ A_{\rm Flat} ^{(l)}$. 
FIG.\ref{areahenka} shows the dependence of 
$\left. A_{\rm EH} ^{(l)}  \right/ A_{\rm Flat} ^{(l)}$ 
on $(m \lambda ^2$,~ $j^2 / m^3)$. 
FIG.\ref{areadannmenoa} and FIG.\ref{areadannmenoc} 
show the behaviors of 
$\left. A_{\rm EH} ^{(l)}  \right/ A_{\rm Flat} ^{(l)}$ 
along 
the two boundaries, line $OA$ and line $OC$, respectively.
FIG.\ref{areadannmenoa} corresponds to the $j=0$ case 
which was discussed in \cite{IKT}.
In the non-rotating case, 
the ratio always satisfies
$\sqrt{2}< \left. A_{\rm EH} ^{(l)}  \right/ A_{\rm Flat} ^{(l)} < 4$.
However in rotating case there is a situation such that 
$0 < \left. A_{\rm EH} ^{(l)}  \right/ A_{\rm Flat} ^{(l)} < \sqrt{2}$
in the region $OSC$ in FIG.\ref{areahenka}.

\section{Summary and Discussion}

We have constructed new charged rotating
multi-black hole solutions
on the Eguchi-Hanson space in the five-dimensional
Einstein-Maxwell system with a Chern-Simons term and
a positive cosmological constant.
These solutions have the mass parameter $m_i$ for each black hole
and the common angular momentum parameter.
In the case of two black holes with $m_1 = m_2 = m$ for simplicity, 
by virtue of the positive cosmological constant,
these solutions  within some region of the parameters $(m,j)$ describe the situation
such that
two rotating black holes with $\rm \rm S^3$ horizon 
coalesce and convert into a single rotating black hole with 
the $\rm \rm S^3 / {\mathbb Z}_2$ horizon.
On the other hand, two-centered Klemm-Sabra solutions
describe the phyiscal situation such that two rotating black holes with ${\rm S}^3$ horizon
coalesce into a single rotating black hole with ${\rm S}^3$ horizon.

We have also discussed the difference in the horizon area 
between our solutions and 
the two-centered Klemm-Sabra solutions.  
We have set the same initial condition in both solutions as follows:
two black holes have the same masses and angular momentum.
In non-rotating case,
the ratio of areas of black hole after coalescence is  
$\sqrt{2} < \left. A_{\rm EH} ^{(l)}  \right/ A_{\rm Flat} ^{(l)} < 4$ \cite{IKT}.
In contrast, for the large angular momentum in the rotating case, there is the region of parameters where 
the ratio of the horizon areas becomes 
$0 < \left. A_{\rm EH} ^{(l)}  \right/ A_{\rm Flat} ^{(l)} < \sqrt 2$.

As mentioned in Introduction, both solutions in this article describe the 
coalescence of black holes by virtue of a positive cosmological constant. 
Nevertheless, in $\lambda \to 0$ limit our results would suggest some information about the coalescence of two rotating supersymmetric black holes on the flat space (BMPV solutions) and on the Eguchi-Hanson space. Therefore, let us discuss the limit $\lambda\to 0$. Two rotating supersymmetric black holes characterized
by the parameters $(m,j)$ with a total horizon area $A^{(e)}$
coalesce into a single rotating supersymmetric  black hole with a horizon area 
$A_{\rm Flat}^{(l)}=\sqrt{(2m^3 - j^2) / (m^3 - j^2)}A_{\rm Flat}^{(e)}$
on the flat space,
while on the Eguchi-Hanson space $A_{\rm EH}^{(l)} = 2 A_{\rm EH}^{(e)}$,
which is independent of parameters $(m,j)$.
If $2/3 < j^2 / m^3 < 1$,
the area of black hole horizon after the coalescence on the Eguchi-Hanson space 
is smaller than that on the flat space, i.e.,  
$0 < \left. A_{\rm EH}^{(l)}  \right/ A_{\rm Flat}^{(l)} < 1 $.

At first sight, the \lq\lq mapping rule" $(2m,~2j)\to(4m,~8j)$ 
for our solutions seems to be 
inconsistent with  the conservation laws of energy and angular momentum. 
Hence, finally let us check the consistency
between the conservation laws 
and the \lq\lq mapping rules" in the $\lambda \to 0 $ limit.
As discussed in the previous section, 
we suppose that 
each black hole on the flat space  
has the same mass, angular momentum and horizon area 
as that on the Eguchi-Hanson space 
at the early time. 
Then, the total mass and angular momentum 
at the early time, 
$M^{(e)}$ and $J^{(e)}$, 
for two black holes  
on the flat space and on the Eguchi-Hanson space  
are given by 
\begin{eqnarray}
 M_{\rm Flat}^{(e)} &=& M_{\rm EH}^{(e)} 
  = 2 \times \frac{3m}{8 \pi G_5} A_{\rm \rm S^3}, 
\\ 
 J_{\rm Flat}^{(e)} &=& J_{\rm EH}^{(e)}   
 = - 2 \times \frac{j}{4 \pi G_5} A_{\rm \rm S^3}, 
\end{eqnarray}
where it is noted that 
$J_{\rm Flat}^{(e)}$ and $J_{\rm EH}^{(e)}$ 
are the angular momenta associated with the Killing vector field $\partial/\partial\psi$.
Of course, 
these amounts are conserved during the processes, 
i.e., 
the total mass and angular momentum
at the late time, 
$M^{(l)}$ and $J^{(l)}$, 
become 
$M^{(l)} = M^{(e)}$ and $J^{(l)} = J^{(e)}$. 
Then 
$M^{(l)}$ and $J^{(l)}$ 
for a single black hole   
on the flat space and on the Eguchi-Hanson space  
are given by 
\begin{eqnarray}
 M_{\rm Flat}^{(l)} &=& M_{\rm EH}^{(l)} 
  = \frac{3m}{4 \pi G_5} A_{\rm \rm S^3},  \label{mfl}
\\ 
 J_{\rm  Flat}^{(l)} &=& J_{\rm EH}^{(l)}   
 = - \frac{j}{2 \pi G_5} A_{\rm \rm S^3}.   \label{jfl}
\end{eqnarray}

On the other hand, 
according to the \lq\lq mapping rule"   
of the two-centered Klemm-Sabra solutions \eqref{2CKS} 
in the $\lambda \to 0 $ limit,   
the mass and the angular momentum  
of a single rotating black hole 
with the parameters $(2m, ~2j)$ after coalescence 
are given by  
\begin{eqnarray}
  M_{\rm Flat}^{(l)} &=& \frac{3 \times 2m}{8 \pi G_5} A_{\rm S^3}
                            =  \frac{3m}{4 \pi G_5} A_{\rm S^3}, \label{mfl1}
\\ 
  J_{\rm Flat}^{(l)} &=& - \frac{2j}{4 \pi G_5} A_{\rm S^3}
       = - \frac{j}{2 \pi G_5} A_{\rm S^3}.  \label{jfl1}
\end{eqnarray}
According to the \lq\lq mapping rule" of our solutions (\ref{EHBH}) 
in the $\lambda \to 0 $ limit, while
the mass and 
the angular momentum 
of a single rotating black hole 
with the parameters $(4m, ~8j)$ after coalescence 
are given by  
\begin{eqnarray}
  M_{\rm EH}^{(l)} &=& \frac{3 \times 4m}{8 \pi G_5} A_{\rm S^3/{\mathbb Z}_2} 
       = \frac{3m}{4 \pi G_5} A_{\rm S^3},  \label{mel1}
\\ 
  J_{\rm EH}^{(l)} &=& - \frac{8j / 2}{4 \pi G_5}\frac{A_{\rm S^3}}{2}
       = - \frac{j}{2 \pi G_5} A_{\rm S^3},  \label{jel1}
\end{eqnarray}
where the factor $1/2$ in $8j/2$ of Eq.(\ref{jel1}) reflects that 
the Killing vector we used to compute the angular momentum is 
$\partial/\partial\psi$ rather than $\partial/\partial(\psi/2)$, 
and the factor $1/2$ in $A_{\rm S^3} / 2$ reflect that the spatial infinity 
is topologically the lens space $\rm S^3/{\mathbb Z}_2$.
Then, we see that 
$M_{\rm Flat}^{(l)} = M_{\rm EH}^{(l)}$ 
from (\ref{mfl1}), (\ref{mel1})    
and 
$J_{\rm Flat}^{(l)} = J_{\rm EH}^{(l)}$   
from (\ref{jfl1}), (\ref{jel1}). 
These relations are same as (\ref{mfl}) and (\ref{jfl}), respectively. 
Thus, the \lq\lq mapping rules" of our solutions (\ref{EHBH}) means only the conservation laws of mass and angular momentum
in the $\lambda \to 0 $ case.

Finally, we mention that
one can generalize our solution (\ref{EHBH}) by replacing the harmonics in 
(\ref{harmH}), (\ref{harmV}) and (\ref{harmome}) by
\begin{eqnarray}
H &=& \lambda \tau + \sum_{i} \frac{M_i}{ \left| \bm{r}-\bm{r}_i \right|},\quad
V^{-1}= \epsilon + \sum_{i} \frac{N_i}{ \left| \bm{r}-\bm{r}_i \right|},
\end{eqnarray}
and
\begin{eqnarray}
{\bm \omega} &=& \sum_{i} 
   \frac{N_i (z-z_i)}{\left| \bm{r}-\bm{r}_i \right|}~
   \frac{(x-x_i) dy -(y-y_i) dx}{(x-x_i)^2+(y-y_i)^2}, 
\end{eqnarray}
respectively. 
Here, the constant $\epsilon $ takes the value $0$ or $1$.  
Black hole solutions on the multi-centered Eguchi-Hanson spaces
are obtained by $\epsilon =0$ with the sum $i\geq 2$.  
Black hole solutions on
the multi-centered-Taub-NUT spaces are obtained by $\epsilon =1$ with the sum $i\geq 1$.
These solutions include some previously known solutions, i.e., 
cosmological non-rotating multi-black hole solutions on the multi-centered-Taub-NUT space 
\cite{IIKMMT}
and rotating multi-black hole solutions on the multi-centered-Taub-NUT space 
with no cosmological constant \cite{GSY}. 
We will study the coalescence of rotating multi-black holes 
on the multi-centered-Taub-NUT space 
in near future.

\section*{Acknowledgments}
We would like to thank Yasunari Kurita, Toshiharu Nakagawa, 
Ken-ichi Nakao and Chul-Moon Yoo  
for useful discussions.
This work is supported by the Grant-in-Aid
for Scientific Research No.13135208 and No.19540305.

\end{document}